\documentclass[twocolumn,prb,amsmath,superscriptaddress,nofootinbib,amssymb]{revtex4}
\usepackage{graphics}
\usepackage{amssymb}
\usepackage{graphicx}
\usepackage{times}
\usepackage{epsfig}          %% um eps-Dateien einzubinden (\epsfig{file=...})

\begin{document}

\title{Single Dirac-cone on the Cs-covered topological insulator surface Sb$_2$Te$_3$(0001)}

\author{Christoph Seibel}\affiliation{Experimentelle Physik VII and R\"ontgen Research Center for Complex Materials (RCCM), Universit\"at W\"urzburg Am Hubland, D-97074 W\"urzburg, Germany}\affiliation{Karlsruhe Institute of Technology KIT, Gemeinschaftslabor f\"ur Nanoanalytik, D-76021 Karlsruhe, Germany}
\author{Henriette Maa{\ss}}\affiliation{Experimentelle Physik VII and R\"ontgen Research Center for Complex Materials (RCCM), Universit\"at W\"urzburg Am Hubland, D-97074 W\"urzburg, Germany}\affiliation{Karlsruhe Institute of Technology KIT, Gemeinschaftslabor f\"ur Nanoanalytik, D-76021 Karlsruhe, Germany}
\author{Minoru Ohtaka}\affiliation{Graduate School of Advanced Integration Science, Chiba University, Chiba 263-8522, Japan}
\author{Sebastian Fiedler}\affiliation{Experimentelle Physik VII and R\"ontgen Research Center for Complex Materials (RCCM), Universit\"at W\"urzburg Am Hubland, D-97074 W\"urzburg, Germany}\affiliation{Karlsruhe Institute of Technology KIT, Gemeinschaftslabor f\"ur Nanoanalytik, D-76021 Karlsruhe, Germany}
\author{Christian J\"unger}\affiliation{Experimentelle Physik VII and R\"ontgen Research Center for Complex Materials (RCCM), Universit\"at W\"urzburg Am Hubland, D-97074 W\"urzburg, Germany}\affiliation{Karlsruhe Institute of Technology KIT, Gemeinschaftslabor f\"ur Nanoanalytik, D-76021 Karlsruhe, Germany}
\author{Chul-Hee~Min}\affiliation{Experimentelle Physik VII and R\"ontgen Research Center for Complex Materials (RCCM), Universit\"at W\"urzburg Am Hubland, D-97074 W\"urzburg, Germany}\affiliation{Karlsruhe Institute of Technology KIT, Gemeinschaftslabor f\"ur Nanoanalytik, D-76021 Karlsruhe, Germany}
\author{Hendrik Bentmann}\affiliation{Experimentelle Physik VII and R\"ontgen Research Center for Complex Materials (RCCM), Universit\"at W\"urzburg Am Hubland, D-97074 W\"urzburg, Germany}\affiliation{Karlsruhe Institute of Technology KIT, Gemeinschaftslabor f\"ur Nanoanalytik, D-76021 Karlsruhe, Germany}
\author{Kazuyuki~Sakamoto}\affiliation{Graduate School of Advanced Integration Science, Chiba University, Chiba 263-8522, Japan}
\author{Friedrich Reinert}\affiliation{Experimentelle Physik VII and R\"ontgen Research Center for Complex Materials (RCCM), Universit\"at W\"urzburg Am Hubland, D-97074 W\"urzburg, Germany}\affiliation{Karlsruhe Institute of Technology KIT, Gemeinschaftslabor f\"ur Nanoanalytik, D-76021 Karlsruhe, Germany}
\date{\today}

\begin{abstract}
Using angle-resolved photoelectron spectroscopy we investigate the surface electronic structure of the three-dimensional topological insulator (TI) Sb$_2$Te$_3$(0001). Our data show the presence of a topological surface state in the bulk energy gap with the Dirac-point located above the Fermi level. The adsorption of Cs-atoms on Sb$_2$Te$_3$(0001) gives rise to a downward energy shift of the electronic valence band states which saturates at a value of $\sim$200~meV. For the saturation coverage the Dirac-point of the linearly dispersive surface state resides in close proximity to the Fermi level. The electronic structure of the Cs/Sb$_2$Te$_3$ interface therefore considerably deviates from previously studied metal-TI interfaces based on the isostructural compound Bi$_2$Se$_3$ which points to the importance of atomic composition in these hetero systems.        
\end{abstract}
\maketitle

Three-dimensional topological insulators (TIs) are currently generating widespread scientific interest in the condensed matter physics community as the distinct topology of their bulk band structure provokes the existence of robust metallic surface states with unique physical properties.\cite{Hasan:10.11,Zhang:11.10} The surface states locally span the global energy gap in the electronic excitation spectrum of the bulk material and their existence is protected by time-reversal symmetry.\cite{Kane:05.9,Zhang:06.3} A salient feature of topological surface states (TSSs) lies in their characteristic spin structure introduced by spin-orbit coupling which locks the spin to the direction perpendicular to the wave vector.\cite{hsieh:09} As a consequence of this spin structure the backscattering of the surface state electrons from non-magnetic impurities is strongly suppressed.\cite{Yazdani:09.8} Currently, most research on TIs is devoted to the chalcogenide semiconductors Bi$_2$Se$_3$\cite{Hasan:09.5,zhang:09.5} and Bi$_2$Te$_3$\cite{Shen:09,zhang:09.5}, related ternary compounds\cite{Shimada:10.9,Yoichi:10.9} as well as to HgTe quantum wells.\cite{Molenkamp:11.3} The TSSs of these materials show a particularly simple dispersion consisting of a single spin-polarized Dirac-cone.

The experimental realization of the most appealing properties of TSSs that have been predicted so far will require interface structures between TIs and metal films. This holds for example for the topological magnetoelectric effect at the interface of a TI and a ferromagnet\cite{Zhang:08.11,Zhang:09.2} as well as for Majorana fermions at the interface of a TI and a superconductor.\cite{Kane:08} It is therefore important to investigate the influence of metallic adlayers on the electronic structure of TI surfaces. Surface-sensitive spectroscopic techniques and in particular angle-resolved photoelectron spectroscopy (ARPES), which also played a key role in the discovery of TIs \cite{Hasan:10.11}, are suitable methods to study modifications in the electronic structure during the formation of interfaces \cite{forster:06}. Indeed, great experimental effort on the basis of ARPES is currently directed towards an improved understanding of the influence of adsorbates on the electronic structure of TI surfaces.\cite{Hasan:11.1,Ast:11.10,Hofmann:11.8,Valla:12.3,rader:12.6,chen:12,Hasan:12.6} However, most of these works have focused on the TI surface Bi$_2$Se$_3$(0001). It therefore appears essential to expand the present investigations to other TI surfaces in order to obtain a broader picture.

The studies mentioned above showed that several adsorbed metal species, such as Cu,\cite{Hasan:12.6} Fe,\cite{Hasan:11.1,rader:12.6} Rb,\cite{Valla:12.3} Cs\cite{Valla:12.3} or Gd\cite{Valla:12.3}, shift the electronic states at the Bi$_2$Se$_3$(0001) surface, including the TSS, to higher binding energies by several hundred meV, possibly due to band bending effects. In all cases the Dirac-point of the TSS is thus located approximately 0.5~eV or even further below the Fermi energy when the saturation coverage is reached for which the electronic states cease to shift upon further increase of coverage (see e.g. Ref.~\onlinecite{Valla:12.3}). Furthermore, significant deviations from the linear dispersion close to the Dirac-point have been observed to occur for the TSS of Bi$_2$Se$_3$(0001) upon metal adsorption.\cite{Valla:12.3,Hasan:11.1} We will show in the following that the situation for the interface Cs/Sb$_2$Te$_3$(0001) is largely different from the mentioned examples involving Bi$_2$Se$_3$.

Despite early theoretical predictions of Sb$_2$Te$_3$ as a 3D TI,\cite{zhang:09.5} experimental investigations of its surface electronic structure have remained scarce when compared to the isostructural counterparts Bi$_2$Se$_3$ and Bi$_2$Te$_3$.\cite{hsieh:09,wang:10} Recent spin-resolved ARPES experiments provide direct evidence for the non-trivial topology of Sb$_2$Te$_3$.\cite{pauly:12} Furthermore, remarkable progress in the fabrication of high-quality MBE-grown samples has recently been reported.\cite{Jiang:12.1,Jiang:12.2}   

In this paper we investigate the electronic structure of the clean Sb$_2$Te$_3$(0001) surface and its modification upon Cs-adsorption using ARPES experiments. For the pristine surface we find the high binding energy tails of the TSS below the Fermi level whereas the Dirac-point is located in the unoccupied states. Our measurements indicate the valence band maximum of Sb$_2$Te$_3$(0001) to be located along the $\bar\Gamma$$\bar M$-direction, significantly off the $\bar\Gamma$-point. The deposition of Cs on Sb$_2$Te$_3$(0001) results in a shift of the electronic states to higher binding energies of up to $\sim$200~meV. The energetic position of the Dirac-point of the TSS is successively moved towards the Fermi energy $E_F$ as the Cs-coverage is increased and finally saturates approximately 65~meV above $E_F$.

The ARPES data were collected in the laboratories in W\"urzburg (Germany) and Chiba (Japan). In W\"urzburg we employed a Gammadata SES 200 electron spectrometer in combination with a monochromatized He discharge lamp operated at an excitation energy of 21.22~eV (He~$\mbox{I}_{\alpha}$). The energy and angular resolution of the setup are approximately $\Delta$\textit{E}\,=\,5\,meV and $\Delta\theta$\,=\,0.3$^\circ$, respectively. In Chiba we used a Scienta R4000 electron analyzer as well as a monochromatized Xe discharge lamp (MB Scientific) operated at an excitation energy of 8.44~eV (Xe~$\mbox{I}_{\alpha}$). The energy and angular resolutions were $\Delta$\textit{E}\,=\,8\,meV and $\Delta\theta$\,=\,0.3$^\circ$. The STM data were obtained in constant current mode using a tungsten tip. All measurements were carried out at room temperature and at base pressures lower than $2\times10^{-10}$ mbar. The preparation of pristine Sb$_2$Te$_3$(0001) surfaces was performed by cleaving commercially available single crystals under ultrahigh vacuum conditions. Cs was deposited at room-temperature using commercial alkali dispensers (SAES getters S.p.A.). The Cs-coverages were estimated on the basis of STM measurements.

\begin{figure}
\includegraphics[width=2.9in]{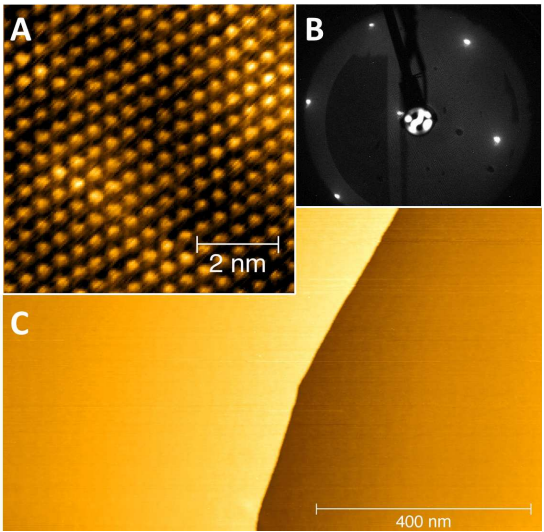}
\caption{\label{fig1}(Color online) Surface characterization of a freshly cleaved Sb$_2$Te$_3$(0001) single crystal.
(a) shows an atomic resolution STM measurement of a 7~nm$\times$~7nm surface area (tunneling parameters $U=-20$~mV and $I=0.1$~nA). The LEED pattern in (b) at 80~eV confirms the hexagonal symmetry of the surface structure. The large area STM scan (\mbox{1~$\mu$m}$\times$\mbox{1~$\mu$m}) in (c) shows two flat terraces seperated by a sharp step edge with a height of $\sim$10~{\AA} corresponding to the height of one quintuple layer. (tunneling parameters $U=-0.4$~V and $I=0.1$~nA).}
\end{figure}

Sb$_2$Te$_3$ crystallizes in the same rhombohedral crystal structure as the TIs Bi$_2$Se$_3$ and Bi$_2$Te$_3$. Along the (0001) direction the crystal structure consists of five-atom layers (quintuple layers) which are weakly coupled via van der Waals-type bonds. An analysis of the surface morphology of the freshly cleaved Sb$_2$Te$_3$(0001) is shown in Fig.~\ref{fig1}. The overview STM image in Fig.~\ref{fig1}.(c) shows that the surface consists of large, flat terraces which are separated by step edges with a height corresponding to one quintuple layer ($\sim$10~{\AA} ). The hexagonal symmetry of the surface unit cell is evidenced by low-energy-electron-diffraction (LEED) in Fig.~\ref{fig1}.(b) and by the atomic-scale STM image in Fig.~\ref{fig1}.(a).   

\begin{figure}[b]
\includegraphics[width=2.9in]{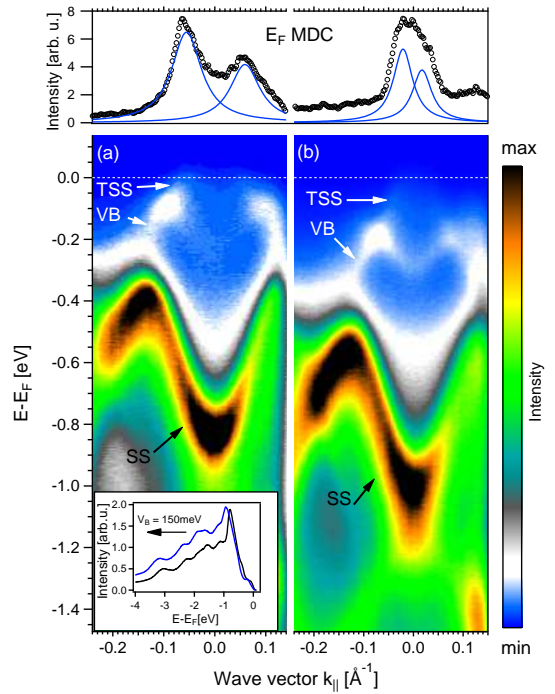}
\caption{\label{fig2}(Color online) Angle-resolved photoemission along the $\bar\Gamma$$\bar K$-direction for clean and Cs-covered Sb$_2$Te$_3$(0001) in (a) and (b), respectively, using He~I$_{\alpha}$-excitation [($\theta_{Cs}=$~0.05~ML)]. The topological surface state (TSS), a valence band state (VB) and another surface state (SS) are labeled in (a) and (b). Momentum distribution curves (MDCs) at the the Fermi level $E_F$ are depicted above the respective angle-resolved spectra. The inset in (a) shows angle-integrated spectra of the complete valence band before (black) and after (blue) Cs-deposition. The adsorption of Cs results in a shift the electronic states to higher binding energies by $V_B =150$~meV.}
\end{figure}

The electronic structure of Sb$_2$Te$_3$(0001) close to the Fermi energy $E_F$ is addressed by ARPES in Fig.~\ref{fig2}.(a). The data shows three dispersive spectral features that are labeled in (a). A linearly dispersing band passes $E_F$ at a Fermi vector of $k_F=0.055(10)${\AA}$^{-1}$. The Fermi-level crossing is also inferred from the momentum distribution curve (MDC) at $E_F$ which is shown above the angle-resolved spectrum in (a). As will be confirmed below, we indentify this feature as the high binding energy part of the Dirac-cone formed by the TSS. At a binding energy of $\sim$110~meV the TSS merges with the valence band state VB. At the connecting points of these states the photoemission intensity is markedly increased. A similar effect has been observed for Bi$_2$Se$_3$(0001).\cite{Hasan:09.5} At higher binding energies we find the prominent feature SS which has very recently been identified as a surface state with Rashba-type spin-splitting by SARPES and DFT-calculations.\cite{pauly:12} The band shows a parabolic dispersion close to the $\bar\Gamma$-point with a positive effective mass of $m^* =0.085 (15) m_e$. At larger wave vectors the band dispersion deviates from the parabolic shape and develops a maximum at $k_{\|}=0.13(1)${\AA}$^{-1}$. 

\begin{figure}[t]
\includegraphics[width=3in]{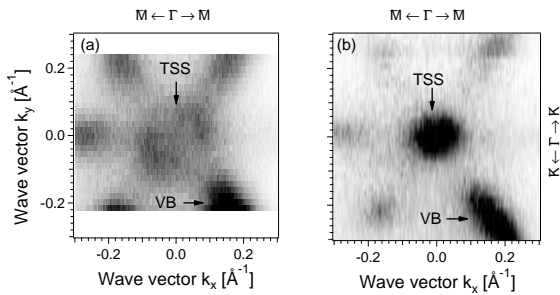}
\caption{\label{fig3} ARPES Fermi surfaces for clean and Cs-covered Sb$_2$Te$_3$(0001) in (a) and (b), respectively [($\theta_{Cs}=$~0.075~ML)]. The data were aquired using Xe~I$_{\alpha}$-excitation. The wave vectors $k_x$ and $k_y$ are oriented along $\bar\Gamma$$\bar M$ and $\bar\Gamma$$\bar K$, respectively. The Fermi surfaces consist of a circular feature around the $\bar\Gamma$-point, arising from the topological surface state (TSS), and additional lobes along the $\bar\Gamma$$\bar M$-directions which are attributed to the valence band states of bulk Sb$_2$Te$_3$. For the Cs-covered surface the size of the circular TSS Fermi surface is contracted indicating that the Dirac-point has been shifted towards $E_F$ [compare Fig.~\ref{fig4}.(b)].}
\end{figure}

Our data in Fig.~\ref{fig2}.(a) indicate that in the vicinity of the $\bar\Gamma$-point the Fermi surface  Sb$_2$Te$_3$(0001) is determined solely by the TSS. This situation is changed at larger wave vectors which becomes apparent by consulting the ARPES Fermi surface map in Fig.~\ref{fig3}.(a). Close to the $\bar\Gamma$-point we find a circular feature which is attributed to the TSS by comparison with Fig.~\ref{fig2}.(a). Away from the center of the surface Brillouin zone we observe additional lobes along the six $\bar\Gamma$$\bar M$-directions which are identified as valence band states of bulk Sb$_2$Te$_3$. This interpretation agrees with previous theoretical predictions and indicates the valence band maximum of Sb$_2$Te$_3$(0001) to be located between the $\bar\Gamma$-point and the $\bar M$-point.\cite{Hasan:09.9,Chulkov:10,zhang:09.5} We conclude that the Fermi level of our samples is determined by intrinsic $p$-type doping and thus is located within the bulk valence band.    

\begin{figure*}
\includegraphics[width=6in]{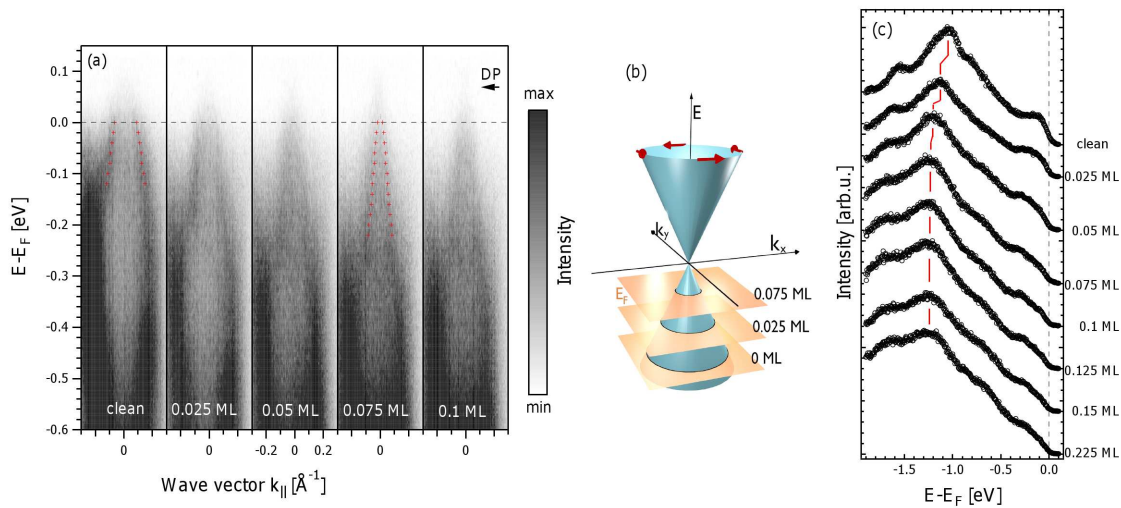}
\caption{\label{fig4}(Color online) Effect of Cs-deposition on the topological surface state of Sb$_2$Te$_3$(0001). The photoemission data in (a) and (c) were acquired using Xe~I$_{\alpha}$ excitation. Angle-resolved spectra for increasing amounts of Cs on the surface are shown in (a). The Dirac-cone dispersion of the topological surface state gradually shifts to lower energies. The data points in (a) for the spectra at coverages of 0~ML and 0.075~ML were obtained from fits to momentum distribution curves (MDCs). The approximate position of the Dirac-point (DP) for the Cs-covered surface is indicated in the spectrum for 0.1~ML. Angle-integrated spectra of the data in (a) and for higher coverages are displayed in (c). Above a coverage of 0.075~ML the electronic states show no further shift to lower energies with increasing Cs-deposition. The schematic in (b) summarizes the experimentally deduced situation for the Cs/Sb$_2$Te$_3$(0001) interface.}
\end{figure*}

Having characterized the pristine surface we now set out to discuss the influence of Cs-adsorption on its electronic structure. In Fig.~\ref{fig2}.(b) we present ARPES data for Sb$_2$Te$_3$(0001) for a Cs-coverage $\theta_{Cs}$ of 0.05~ML. The changes brought about by the adsorption are inferred by comparison with the spectrum of the clean surface in Fig.~\ref{fig2}.(a). We find an overall downshift in energy of the electronic states as is most clearly seen for the parabolic surface state SS. Also the energies of the valence band feature VB and of the TSS are lowered after the adsorption. The Fermi wave vector of the TSS is reduced to $k_F=0.020(5)${\AA}$^{-1}$ suggesting that the Dirac-point, which still lies in the unoccupied states, approaches the Fermi level. The reduction of $k_F$ due to Cs-adsorption is also clearly inferred by comparing the MDCs at $E_F$ in Fig.\ref{fig2}.(a) and (b). In agreement with previous results our data in Fig.~\ref{fig2}.(b) shows that Cs-deposition provides a means to change the position of the Dirac-point relative to $E_F$.\cite{wang:10} An overview on the influence of the Cs-adsorption on the valence band states of Sb$_2$Te$_3$ is gained from the inset in Fig.~\ref{fig2}.(a) which shows angle-integrated data over a large energy interval. The spectra evidence an energy shift $V_B=150$~meV of the entire valence band suggesting that the Cs-adsorption induces a downward band bending. Adsorbate-induced band bending on semiconductor surfaces is a well-known phenomenon\cite{Moench:95} which recently has also been observed for other TI surfaces, such as in particular Bi$_2$Se$_3$(0001).\cite{Hofmann:11.8,Hsieh:09.7,bianchi:10.11,Noh:08} Notice, however, that the energy shift of the electronic band structure is not entirely rigid as is inferred from the energetic position of the connecting point of the features TSS and VB in Fig.~\ref{fig2}. This point in the electronic band structure is changed by only $\sim$90~meV; a value that is significantly smaller than $V_B$. This observation suggests that other surface effects may play an additional role and that a simple rigid band bending scenario does not fully capture the experimental results. Changes in the dispersion of surface related electronic features as a result of alkali adsorption have, for instance, also been observed in metallic surface systems.\cite{bentmann:09}     

In order to analyze the effect of Cs-adsorption on the TSS in more detail we consider the coverage dependent data in Fig.~\ref{fig4}. Panel (a) shows ARPES spectra close to $E_F$ for increasing amounts of Cs on the surface. The dispersion of the TSS is successively shifted to lower energies as $\theta_{Cs}$ is increased up to 0.075~ML. In accordance with the Dirac-cone shape of the dispersion of the TSS, the energy shifts are accompanied by a gradual decrease of the Fermi vector. This effect is also nicely visible by comparing the two Fermi surface (FS) maps for the clean and the Cs-covered surface in Fig.~\ref{fig3} where a contraction of the circular FS of the TSS as a result of the adsorption is revealed. For $\theta_{Cs}=0.075$~ML the full Dirac-cone dispersion is visible in the ARPES spectrum in Fig.~\ref{fig4}.(a). Notice that due to thermal occupation ARPES gives access to states up to $5k_B T$ above the Fermi level which amounts to $\sim$125~meV in the present case (room temperature).\cite{Greber:97} The Dirac-point is located slightly above the Fermi energy at $E_D =(65\pm 20)$~meV. Note that the dispersion of the TSS is remarkably linear which agrees with theorectical predictions for the clean surface.\cite{pauly:12,Chulkov:10} The high degree of linearity in the dispersion of TSS of Sb$_2$Te$_3$(0001) has also been emphasized based on the results of Landau level spectroscopy.\cite{Jiang:12.1} In the present case it is also important to note that the linearity of the dispersion is not affected by the Cs-adsorption. A summary of the experimental findings is provided by the schematic in Fig.~\ref{fig4}.(b).

Upon further increase of $\theta_{Cs}$ we find no additional changes in the energies of the electronic states and in particular $E_D$ remains constant. This saturation behavior of the Cs-induced energy shifts is also clearly seen in the integrated spectra in panel (c). Up to a coverage of $\sim$0.075~ML the energies of the electronic states are lowered but above this value they remain constant, indicating a maximal band bending of $\sim$200~meV. 

It is instructive to compare our results with previous findings for the TI surface Bi$_2$Se$_3$(0001). Similarly to the present case, metal adsorption on this surface was observed to induce a downward shift of the electronic states to lower energies.\cite{Hasan:12.6,Hasan:11.1,rader:12.6,Valla:12.3} Unlike for Sb$_2$Te$_3$, however, the energetic position of the Dirac-point at the respective saturation coverages was determined to lie far below the Fermi level for interfaces with Bi$_2$Se$_3$(0001) ($\sim$500~meV), even in the case of bulk $p$-doped samples\cite{Hasan:12.6,Hasan:11.1}. The presence of metal atoms on Bi$_2$Se$_3$(0001) was also found to strongly affect the dispersion of the TSS close to the Dirac-point giving rise to significant deviations from its linearly dispersive behavior,\cite{Valla:12.3} an effect that is not observed in the present case. Both of these latter observations may constitute major obstacles as far as probing the properties of Dirac-fermions at interfaces of the TI Bi$_2$Se$_3$ with magnetic or non-magnetic metal species is concerned. Our results on Sb$_2$Te$_3$ suggest that, in order to overcome these issues, it can be useful to modifiy the atomic composition of the studied TI materials because the adsorbate-induced modifications in the surface electronic structure can vary significantly depending on the specific atomic constituents.    

In summary, we have investigated the metal-TI interface Cs/Sb$_2$Te$_3$(0001) by ARPES measurements. Due to intrinsic $p$-doping of the bulk crystals the Dirac-point of the topological surface state on the pristine surface is located significantly above the Fermi level, whereas its high-binding energy tails extend into the occupied regime. Confirming previous theoretical results we find the valence band maximum of Sb$_2$Te$_3$(0001) to reside along the $\bar\Gamma$$\bar M$-direction of the surface Brillouin zone. The adsorption of Cs is observed to induce coverage dependent shifts of the valence band states to higher binding energies which saturate at $\sim$200~meV. These energetic changes of the valence states are accompanied by shifts of the Dirac-cone dispersion of the topological surface state. For the saturation coverage the Dirac-point of the topological surface state is located $\sim$65~meV above the Fermi energy.      

\section{Acknowledgements}
H.~M. would like to thank Takashi Aoki, Kenta Morioka and Hirotaka Ishikawa for experimental assistance. This work was supported by the Bundesministerium f\"ur Bildung und Forschung (Grants No. 05K10WW1/2 and No. 05KS1WMB/1), the Deutsche Forschunsgsgemeinschaft within the Forschergruppe 1162 (P3) and the Grant-in-Aid for Scientific Research (A) 20244045.
\bibliographystyle{apsrev}

\end{document}